\documentclass[12pt]{article}
\usepackage{a4} 
\usepackage{epsfig}
\begin{document}
\title{Diffusion in a strongly correlated anisotropic overlayer}
\author{Igor F. Lyuksyutov$^{a}$\footnote{Also at Institute of Physics, 252028 Kiev, Ukraine},
H.-U. Everts $^{b}$
 and  H. Pfn\"ur$^c$\footnote{corresponding author, e-mail: pfnuer@fkp.uni-hannover.de}\\
$^{a}$Department of Physics, Texas A\&M University,\\
College Station, TX 77843-4242,  USA \\
 $^{b}$Institut f\"ur Theoretische Physik, Universit\"at Hannover, and \\
$^c$Institut f\"ur Festk\"orperphysik, Universit\"at Hannover,\\
Appelstr. 2, D-30167, Hannover, Germany\\[1cm] 
} 
\date{\today}
\maketitle
\makeatletter
KEYWORDS: Surface diffusion, surface defects,  
equilibrium and non-equilibrium thermodynamics and statistical mechanics. 

\makeatother
\begin{abstract}
We study the collective diffusion in 
chain structures on anisotropic substrates
like (112) bcc and (110) fcc surfaces
with deep troughs in the substrate potential corrugation.
These chain structures are aligned normal to 
the troughs and can move only along the troughs.
In a combination of theoretical arguments and of numerical simulations, 
we study the mass transport in these
anisotropic systems.  
We find that a mechanism similar to soliton diffusion, 
instead of single particle diffusion, is still effective  at temperatures
well {\it above} the  melting temperature of the ordered chain structures.
This mechanism is directly correlated with the ordered phases 
that appear at much lower temperatures.
As a consequence, also the influence of frozen
disorder is still visible above the melting temperature.
Theoretically we predict a strong dependence of the 
pre-exponential factor and weak dependence 
of the activation energy on the 
concentration of frozen surface defects. These predictions are confirmed 
by the simulations.
\end{abstract}

\section{Introduction}
A key of understanding chemical reaction 
kinetics on surfaces is the 
understanding of diffusion of particles on surfaces. 
Surface diffusion also determines many other surface 
properties, e.g. in adsorbed films acting as lubricants. 
Nevertheless, our present understanding of surface diffusion 
is far from being complete. In particular, 
situations of highly anisotropic diffusion and the role 
of defects is far from being understood. 
Surface diffusion is usually described in terms 
of single atom diffusion (see, 
e.g. \cite{gomer,naum84,naumved}). Especially 
for incommensurate layers, however, this description 
is not adequate for mass transfer by diffusion and for
the very strong influence of frozen defects in this case
(see \cite{naum84,naumved}). 
In order to describe this situation properly, a collective 
``soliton diffusion'' mechanism was proposed 
(see  \cite{lyupok,lyuk86,twod}). 
We show below that this approach can also be
adapted to describe diffusion in chain structures 
on an anisotropic substrate.

\begin{figure}
\epsfxsize=4.truein
\centerline{\epsffile{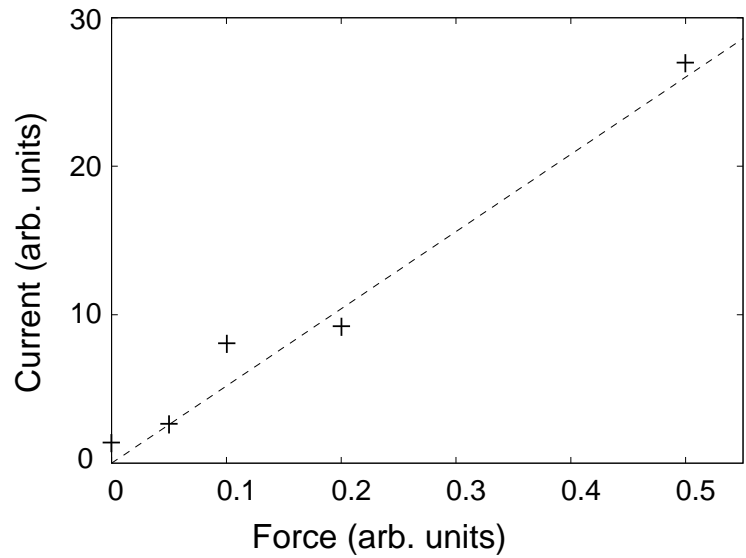}}
\caption{
Diffusion current as a function of applied field (in units $J_0/2$)
\label{f1}}
\end{figure}

Two-dimensional structures
which are formed on surfaces with  strongly
anisotropic substrate potential corrugation 
like (112) bcc and (110) fcc surfaces
have been studied by many authors (for a review, see \cite{med}).
 Chain structures of the type $(1\times p)$
are common on these substrates.
At coverage less than 1/2 monolayer, these 
structures  consist of parallel chains of adsorbed atoms.
A typical example is
schematically shown in Fig.~\ref{f1}. In experiment $p$
can be as large as 8\cite{twod,med}.
The potential corrugation of the (112) bcc surface
can be represented as troughs in the $\langle 111 \rangle$ direction.
The substrate potential is also modulated along the troughs with
the substrate period.
In a typical $(1\times p)$  structure,  chains are oriented
normal to the trough direction.
For the very existence of a chain structure with a large period,
the interaction along the chain $J_0$ should be much larger than
the interaction between chains.

Chain structures demonstrate remarkable thermal
stability (see \cite{twod,med}).
Their transition temperatures are high,
compared with 
the weak interaction between chains in the structures
with large period.
Though chain structures are known for a long time, the
first studies of diffusion in these structures
have appeared only  recently. As a very interesting phenomenon, these experiments
on surface diffusion in pure Li and Sr overlayers
on the Mo(112) face   \cite{lith,lidy,five},
confirm the extremely strong
anisotropy of the diffusion coefficients ($10^3$ times and more)
and reveal a strong correlation between
surface diffusion parameters and the equilibrium phases of the overlayer at temperatures much
higher than the melting temperatures of these phases.
In these experiments the
macroscopic flow of atoms in the overlayer
was actually measured, not the diffusion of single atoms.
A typical system is Li on Mo(112). At low coverages
it forms two chain structures  $(1\times 4)$ at the coverage
 $\theta =.25$ and  $(1\times 2)$ at  $\theta =.5$. At higher
coverages incommensurate structures appear.

Only recently \cite{lidy} an influence of frozen point
disorder (Dy adatoms) on diffusion of Li on Mo(112)
at coverages $1/4 <\theta < 1/2$, which correspond 
to commensurate structures (at low temperatures),
has been reported 
for a Dy coverage $\theta_{Dy}=0.1$. 
As expected, the diffusivity is reduced, 
but mainly by a  drastic decrease of the pre-exponential 
factor $D_0$ (up $10^3$ times), not by changes in the activation energies.

A strong correlation between the low-temperature
phases and the diffusion coefficient at high temperature
has been found experimentally in many strongly interacting 
and strongly anisotropic overlayer systems \cite{naum84,naumved,lyuk86,twod,lith}. 
The pronounced influence of defects, on the
contrary, has previously only been observed 
for incommensurate systems \cite{naum84,naumved,lyuk86,twod,vedula,kleint}.

Motivated by these experimental findings, we study in this work
theoretically the diffusion in chain structures on anisotropic substrates,
and focus on an explanation for the  
correlation between  diffusion 
and the ordered low-temperature structures and on the  
influence of defects. In experiment, these
phenomena have been observed for diffusion along the troughs.
For this reason we focus on diffusion in the same direction,  
although the atom motion in the simulations 
described below is allowed in both directions.
 
Since our study only intends to clarify the basic mechanism of diffusion in this 
class of systems, the details of interactions are not important at this point. 
In fact, we are interested in mass transfer on a scale larger than the 
correlation length
along the chain. 
As will be seen below, there are clear indications that the soliton diffusion mechanism is 
indeed still effective at temperatures far above melting. 
Especially for this mechanism, due to its collective nature, 
the details of the microscopic model should not change
qualitatively our results. For this 
reason, we formulate a simple model which preserves the most important 
properties - strong correlations along the chains and weak 
interaction between the chains. 
%
Since in experiment close relations between the 
ordered structures and diffusion properties 
have been found, we first briefly review in 
Sec. \ref{mass} the equilibrium properties 
of commensurate and incommensurate 
structures on these anisotropic surfaces  (see, e.g., ref.\cite{twod} and references therein) 
before addressing 
possible mechanisms of mass transport. 
We show that indeed soliton diffusion plays an important role on these surfaces, 
and it is this mechanism which makes the 
connection between ordered structures and diffusion at 
much higher temperatures than the melting 
temperatures understandable.
In Sec. \ref{mc} we show that the results of Monte-Carlo simulations
of mass transfer, using a simple anisotropic lattice gas model,  corroborate most 
of the predictions obtained analytically. 

\section{Mass Transport in a Chain Structure.}
\label{mass}

\subsection{Thermodynamics of the  Chain Structure.}
\label{term} 
\begin{figure}
\epsfysize=3.truein
\centerline{\epsffile{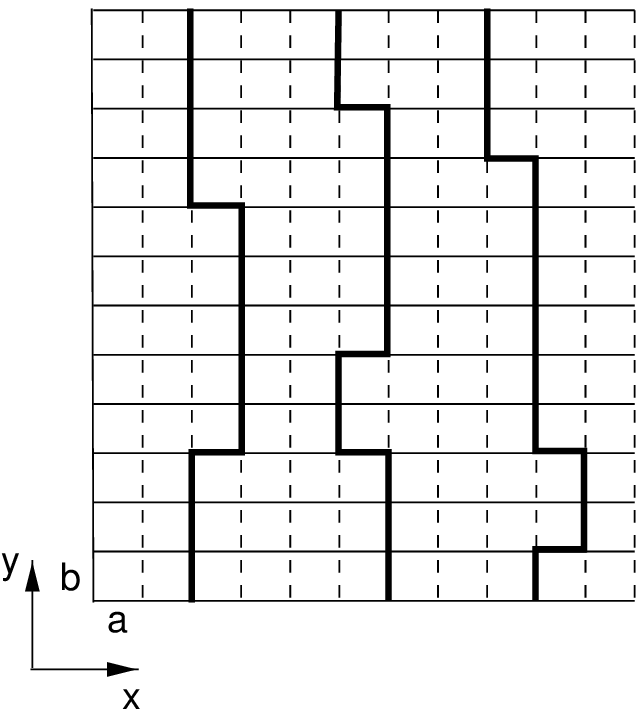}}
\vspace{0.5cm}
\caption{
Schematic representation of the chain structure with kinks. 
Solid lines represent chains of atoms,
thin solid lines represent the trough direction.
Dashed lines represent substrate potential minima. 
\label{f0}}
\end{figure}

Consider first the thermal fluctuations of a single chain
of atoms aligned along the $OY$ axis with period $b$. 
The chain is placed into a periodic
potential with period $a$.
The chain can move from one substrate potential minimum 
to another by creating kinks as shown in Fig.~\ref{f0}.
If $E_k$ is the kink energy, then the mean distance 
between kinks is $b\exp(E_k/T)$. The mean square 
transverse displacement of a chain of length $L$
is:
\begin{equation}
\label{L}
\langle (u(L)-u(0))^2\rangle = \frac{2a^2L}{b\exp(E_k/T)} 
\end{equation} 
Using Eq.~\ref{L} and the corresponding relation for a continuous string, 
one can substitute in the thermodynamic limit the chain of atoms
by a continuous string with the Hamiltonian 
\begin{equation}
\label{Hs}
{\cal H}_s= \frac{\varepsilon}{2}\int dy (\frac{\partial u}{\partial y})^2 
\end{equation} 
where $\varepsilon = (Tb/2a^2)\exp(E_k/T)$ is the  rigidity.

The structure  $(1\times p)$ can occupy $p$ different sub-lattices,
and has $Z_p$ symmetry (the group of additions 
modulo $p$). The most studied models with this symmetry
are the $Z_p$-model, the clock model, and the discrete Gaussian
model (for a review see e.g.\cite{savit}). 
The thermodynamics of the clock model has been studied
in detail (see e.g.\cite{savit}). For $p>4$ three 
phases were shown to exist in the model:
long range
order at $0<T<T_p$,  algebraic decay of correlations at
 $T_p<T<T_m$, and a disordered phase for $T>T_m$, where $T_m$ is the melting 
temperature  and $T_p$ the depinning temperature.

Above  $T_p$ the correlation properties
of the chain structure are the same as those of a lattice of parallel
strings. On scales much larger than the average distance between 
kinks, this lattice can be described by the Hamiltonian: 
\begin{equation}
\label{Hs2}
{\cal H}_{ch}= \frac{1}{2}\int dx dy(K_1(\frac{\partial u}{\partial x})^2 +
K_2(\frac{\partial u}{\partial y})^2 )
\end{equation} 
where 
\begin{equation}
\label{tilt}
K_2 =\frac {\varepsilon}{l}
\end{equation} 
 is the shear modulus.
We assume repulsive interactions between chains, which
decay as a power of the structure period $l=pa$.
The compression modulus $K_1 $ can be
represented as a sum of two terms:

\begin{equation}
\label{comp}
K_1= \frac{J_{\perp}}{ab}\frac{1}{p^{\alpha}}
+\frac{{\pi}^2T^2}{\varepsilon a^3p^3}
\end{equation} 
The first term in Eq.~\ref{comp} is due to 
direct repulsive interaction between chains,
and the second is due to entropic repulsion (see e.g.\cite{twod}).
In the simplest case of the repulsive interactions
between adatoms, which decay as $r^{-\beta}$, 
the exponent $\alpha = \beta$.
For dipole-dipole repulsion or for repulsion due to substrate 
elastic deformations  $\beta =3$ (i.e. $\alpha =3$).
In the case of an anisotropic Fermi surface, for some range 
of distances,  the value of $\beta =1$ \cite{med}.
 $J_{\perp}p^{\alpha +2}$ is of the order of the energy change 
per atom in the chain 
due to the displacement of the chain
by one lattice constant $a$ along the trough.

The melting temperature $T_m$ is given by:
\begin{equation}
\label{Tme}
T_m= \frac{\pi p^2a^2}{8}\sqrt{K_1K_2}
\end{equation} 

Using Eqs.~\ref{tilt} and \ref{comp} 
\begin{equation}
\label{Tm}
T_m\approx \frac{E_k}{\ln{(E_k/J_1)}}
\end{equation} 
follows, where $J_1 = const.\times J_{\perp}$.
In other words, for highly anisotropic chain structures the
melting temperature is defined mainly 
by the kink energy $E_k$. 

For the depinning temperature $T_p$ we have 
the exact relation \cite{pok73}:
\begin{equation}
\label{Tp}
T_p = (16/p^2) T_m
\end{equation} 
When $p=4$ we have  $T_p = T_m$ and there is only one transition 
for $p < 4$. 
Above $T_p$ the thermal fluctuations "smear" the chain lattice
over many potential minima of the substrate potential.
The kink free energy is zero. As a result, 
the chains can move due
to the movement of kinks which can appear simply from thermal fluctuations at 
these temperatures.

Above $T_m$ the system is in the liquid state. This liquid is highly 
anisotropic: the correlation radius $r_c$ is 
large along the chains: $r_{c\vert\vert} \gg b$, 
but small perpendicular to the chains  $r_{c\perp}\approx pa$.

\subsection
{Frozen Defects.}
\label{deff}
The influence of  non-equilibrium 
(frozen) defects  of different kinds 
on the order and phase transitions has been studied 
both theoretically and experimentally 
for a long time in connection with 
magnetic systems (for review see, e.g., ref.\cite{Vill}) 
and particularly with the  spin glass problem
(for review see, e.g.,\cite{Binder}). 
The disorder was  classified as random bond-like
and random field-like. The random bond-like 
disorder interacts with the same strength with
different domains of the pure system. On the 
contrary, the random field-like disorder 
distinguishes between different domains. 
Random bond-like disorder
destroys the order only at some finite concentration.
In the ordered phase random bond-like disorder
changes the critical behavior.
The lower the dimensionality of the system, the stronger are
the effects of frozen defects.
It is generally accepted
that in two-dimensional systems 
the random field-like disorder destroys the long range order at
arbitrarily small values of the random field 
strength even at T=0 (see e.g.\cite{Vill}). 
However, the mean size of the well ordered domains 
(i.e. the correlation radius)  
diverges exponentially when the disorder strength decreases, e.g. in the 
case of the two-dimensional random field Ising model (ref.\cite{Vill}).

In reality no surface, regardless of how carefully it has been prepared,
is free of defects, which are of different origin and dimensionality (point 
or linear). The list includes impurities segregated onto 
the surface, adsorbed impurities, excess atoms of the
crystal proper, steps, etc. At low temperatures, the mobility 
of these defects is extremely small, and they can be treated as
frozen. 
These non-equilibrium defects give rise to various random
fields on the crystalline surface. 

The frozen point defects on a surface, depending on their position 
with respect to substrate and overlayer, correspond
to random bonds in magnetic models (points with 
special symmetry)
or, in general, to both  random field 
and random bonds simultaneously. 

It is generally accepted that the random field defects
destroy order in the system with algebraically 
decaying correlation in spatial dimensions $D \geq 2$.
This result was first obtained by Imry and Ma \cite{ima}.

\subsection
{Mass Transfer Mechanisms.}
\label{difm}

Frozen disorder strongly affects the large scale dynamics of an 
overlayer. In particular, it can lead to slowing down the relaxation 
processes in the layer.
The problem is far 
from being understood, but at least two 
different relaxation
mechanisms are conceivable: (i) via elastic deformation 
of the chain lattice (at $T>T_p$) 
and (ii) via  defect (dislocation) formation 
in the chain lattice in the overlayer (plastic flow).
For the case of the incommensurate overlayer 
the latter mechanism, the soliton diffusion mechanism,
 has been proposed theoretically 
in order to describe diffusion 
experiments in overlayers (see ref.\cite{lyupok} and ref.\cite{twod} Ch.10)
Experiments from 
different experimental groups and for different systems
(refs.\cite{vedula}, \cite{kleint} and ref.\cite{twod} Ch.10) have shown the relevance 
of the soliton diffusion mechanism.
We will apply the approach developed in refs.\cite{lyupok},\cite{twod}
to analyze the mass transfer in chain structures.

\paragraph*{Ideal substrate, $T < T_p$.} 
Below $T_p$ the chain can shift by kink diffusion.  The activation
energy $E_a$ in this case is equal to the sum of
the energy needed to create 
a pair of kinks  $E_a=2E_k$ and of the energy 
barrier $E_{\vert\vert}$ for atom
movement along the trough.
The effective diffusion coefficient can be written in the form:

\begin{equation}
\label{difk}
 D_{eff} \approx \omega_0 b^2\exp(-2E_k/T)\exp(-E_{\vert\vert}/T)=
D_0\exp(-E_{a}/T),
\end{equation} 
The characteristic attempt frequency  $\omega_0$ 
is of the order of the Debye frequency
of the overlayer $\omega_D$. If only nearest neighbor interactions are
considered the activation barrier for
kink diffusion at  $T < T_p$ is close to the activation
barrier for single atom diffusion.

\paragraph*{Ideal substrate, $ T_p <T < T_m$.}
On the ideal substrate the depinning transition should result 
in a drastic change of the apparent activation energy of diffusion.
As we have already mentioned, the kink free energy of the chain 
structure disappears above $T_p$ due
to thermal fluctuations. This implies that in the limit of an infinitely
long observation time the correlation properties of the chain lattice 
are the same as for elastic media. 
However, for finite times the lattice of chains needs to overcome 
activation barriers.
As a result 
in the limit of  infinitely small gradients 
(i.e. long observation time)
the  apparent activation energy
should change from  $2E_k+ E_{\vert\vert}$ to  $E_{\vert\vert}$.
We are not aware of experimental observations of this
situation.

\paragraph*{Substrate with disorder,  $ T_p <T < T_m$.} 
A real substrate always has some amount of defects. 
We have already mentioned
that frozen defects destroy the order at large scale. 
These defects can effectively
pin the chain structure. 
The simplest defect is a point trap. 
We discuss the case when the depth of the trap 
is larger than the interaction between atoms in the chain.
The nature of the trap
and the depth of the potential well may vary. However,
the energy barrier for depinning from traps 
of different depths will not vary very much. 
The main reason for this is the mechanism 
of depinning. E.g. for the situation described the chain can be depinned 
more easily by disrupting itself 
rather than by removing the atom in the trap. This means that the upper
bound on the depinning energy is given by the energy needed 
to break the chain. 

Consider now in more detail the mass transfer above $T_p$.
The gradient of the 
surface density of adatoms induces a
pressure gradient in the depinned chain lattice, which in turn
generates effective deformations at chain pinning sites.
Depinning of the chain from a defect is followed by a displacement
by $l=pa$ of all chains in a region of characteristic
size $a/\sqrt{\theta_d}$, where $\theta_d$ is the defect (trap)
concentration per unit cell. Such a region can be treated as
a particle moving under a pressure-gradient force $f$
that is proportional to the pressure difference $\Delta P$
and to its transverse size:

\begin{equation}
\label{dif1}
f\approx \Delta P \frac{a}{\sqrt{\theta_d}};
\,\,\,\,\,\,\,\,\,
\Delta P = -K_1\frac{\Delta l}{l},
\end{equation} 
where $\Delta l$ is the average change of the structure period
due to compression,  
and the effective compression modulus is given by Eq.~\ref{comp}.
The average  change of period, $\Delta l$, can be evaluated from the
coverage gradient ${\partial \theta}/{\partial x}$ in a region of the
order of the distance between the defects.
As a result we obtain:

\begin{equation}
\label{dif4}
\frac{\Delta l}{l}\approx
\frac{\partial l}{\partial \theta} \frac{\partial \theta}{\partial x} \frac{a}{\sqrt{\theta_d}}
= -\frac{l}{\sqrt{\theta_d}}\frac{\partial \theta}{\partial x}.
\end{equation} 

we have used here the fact that $l=a/\theta$. Finally we obtain for the force:

\begin{equation}
\label{dif2}
f\approx  \frac{K_1 la }{\theta_d}\frac{\partial \theta}{\partial x}
\end{equation} 
The conventional equation for the diffusion flow under a 
constant force (see e.g. \cite{LL}) results in 
the following expression for $D_0$: 
\begin{equation}
\label{dif3}
D_0\approx \omega_0 l^2\frac{K_1 la }{T\theta_d}
\end{equation}

\paragraph*{Mass transfer in the liquid ($T > T_m$).}
At  $T > T_m$ there is no order at large scale.
Correlations along the troughs are short ranged, the 
correlation radius in this direction is not larger
than a few substrate periods.
On the other hand, the correlations along the chains
can remain strong at temperatures far above $T_m$ 
with the correlation radius exceeding many substrate periods.
These latter correlations will significantly influence
the diffusion above  $T_m$
so that one expects to find a close 
connection between the effective diffusion coefficient above  $T_m$ 
and the equilibrium structures at  $T < T_m$.   
At  $T > T_m$  the overlayer  can be considered 
as consisting of anisotropic ``particles'' with
a length along the chains many times larger than the length along
the troughs. 
Let us consider a single chain in such a ``particle''. 
In the thermodynamic limit, i.e. for infinite observation times,
the kink free energy is zero.  As a result
one has a finite single kink density.
However, we discuss the system dynamics for times of the order of
the relaxation times on an  atomic scale. This means that kinks need
to overcome local barriers to move from one point 
on the chain to another.
As soon as a kink appears on the chain,
e.g. at the end of the chain, 
it propagates  along the chain. 
Without defects the barrier it should overcome
is the potential barrier along the trough $E_{\parallel}$.
As a result of kink motion, the chain moves 
by one substrate period along the troughs.
In the case of a  substrate with frozen defects the kink
stops at the frozen defect. The chain will
remain at the same place until the kink overcomes the defect.
As a result, in the case of  strong pinning by defects,
the main contribution to the diffusion time is 
the residence time $\tau_{dep}$ on the defect. One can represent the
"particle" movement by a series of jumps of the order
of the average distance between defects i.e. 
of  $a/\sqrt{\theta_d}$ during the time
interval defined by the residence time  $\tau_{dep}$ on the defect.  
As a result, the effective diffusion coefficient scales with
defect concentration as 

\begin{equation}
\label{dif5}
 D_{eff} \approx  (\frac{a}{\sqrt{\theta_d}})^2 \frac{1}{{\tau_{dep}}}
 \approx\frac{1}{{\theta_d}}\omega_0 a^2\exp(-E_a/T).
\end{equation} 
This result is valid as long as the correlation length along
the chain is larger than the distance between defects.
 Eq.~\ref{dif5} assumes that the residence time  $\tau_{dep}$ is much
larger than the diffusion time between defects  $\tau_{dif}({\theta_d})$.
Both conditions mean that the diffusion behavior reduces to that found on
a surface without defects once the concentration 
of defects falls below a certain
value (which depends on temperature).
This directly avoids  the divergence of $ D_{eff}$
in Eq.~\ref{dif5}, i.e. Eq.~\ref{dif5} considers only
the case when diffusion is fully limited by defects.

\section
{Monte-Carlo Simulations.}
\label{mc}

\subsection
{The Model.}
\label{mod}
In order to study the behavior of the {\it effective} diffusion
coefficient, we use a procedure which simulates in a
most efficient  way  the experimental situation. 
We will not consider diffusive motion
of single particles. Instead, we study the particle current
as a function of applied force. There is no force
applied to each particle in a real experiment. However,
there is a density  gradient resulting in the gradient
of the chemical potential. The chemical 
potential gradient changes very
slowly on the atomic scale.  This gradient 
plays the role of some average force which acts on 
the particles.

The simplest  model of diffusion in a strongly interacting 
overlayer is the two-dimensional lattice gas Ising model 
with Kawasaki dynamics. 
The Hamiltonian of the Ising lattice gas  under
the applied field $F$ in  $x$ direction has the form:
\begin{equation}
\label{H} 
 {\cal H}_b = - \sum_{\bf r, a}
 J_{\bf a}({\bf r}) n_{\bf r}n_{\bf r+a}
 -   \sum_{\bf r}F {x \over \vert{\bf a}\vert} n_{\bf r} 
\end{equation} 
where summation over ${\bf r} = (x,y)$ runs 
over the lattice sites,
 ${\bf a}$ labels nearest 
neighbors.
 Here  $ n_{\bf r} =0,1$ and $J_{\bf a}({\bf r})$ is measured in units 
of $J_0>0$. 

\begin{figure}
\epsfxsize=4.truein
\centerline{\epsffile{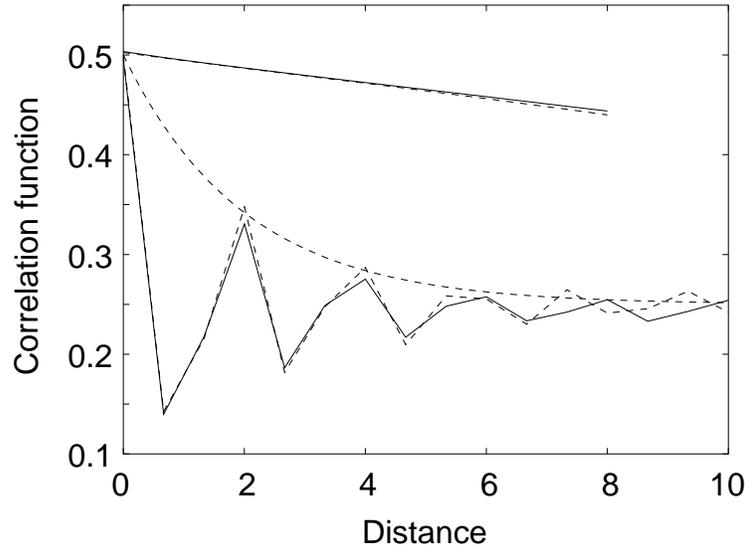}}
\vspace{0.5cm}
\caption{
Particle correlation function in the $24\times 192$ system at a coverage of $\theta = 1/2$ 
after  $2\times 10^6$ MC steps at $T =  4.765 J_0 = 1.23T_{m,1/2}$ with (solid lines) and 
without (dashed lines) external force. Upper curves: correlation along the chains. 
Lower curves: correlation normal to the chains (including an exponential fit to the amplitudes). 
\label{f11}}
\end{figure}

We have used  Monte-Carlo simulations with 
Kawasaki dynamics (see e.g. \cite{bin}) to study 
diffusion. To model the chain structure of the experimental
systems \cite{med} we have introduced a strong attractive 
nearest neighbor interaction ($N$) 
in the $Y$-direction, $J_y = 8 J_0$, and repulsive nearest ($N$) 
and  next nearest ($NN$) neighbor interactions in the $X$-direction, $J_{N} = -2 J_0$ 
and $J_{NN} = - J_0$.
In order to simulate scenarios that are 
close to the experimental situation\cite{naum84,naumved,lyuk86},
we have measured the particle current under the applied force $F$
at temperatures well above the melting 
temperature. First we demonstrate that the particle current is linear in
 $F$. Fig.~\ref{f1} shows this dependence of the current on
 $F$ for $F \leq .25 J_0$ at the lowest temperature used
in our simulations. The linear dependence has been
checked up to  $F \leq  J_0$ at higher temperatures.
We have also checked that the applied force has no
measurable influence on the order in the system.
Fig.~\ref{f11} shows the  
correlation function along the chains and normal to the chains
in a $24\times 192$ system  for the  coverage  $\theta = 1/2$ 
after  $2\times 10^6$ MC steps at $T =  4.765 J_0$. The two different sets
of data correspond to the system with and without applied
force. The data in this figure also directly demonstrate the remaining strong correlations above $T_m$,
in particular along the chains. Even perpendicular to the chains short range order 
over several lattice constants still exists at this temperature, which is roughly 20\% above 
$T_m$.

\subsection
{Diffusion and Equilibrium Phases.}
\label{dif}
At equilibrium the simulated system studied here has three 
ordered structures at low temperatures:
 $(1\times 2)$ (near a coverage $\theta= 1/2$), 
 $(1\times 3)-2X$ (near a coverage $\theta= 2/3$), and
 $(1\times 3)$ (near a coverage $\theta= 1/3$).
 The melting temperature for the
 $(1\times 2)$ structure is $T_{m,1/2} =3.86 J_0 $. Those for the
 $(1\times 3)-2X$ and  $(1\times 3)$ structures are of the same order of magnitude.
As mentioned in Sec. II, there is no depinning in these cases, only melting. 

\begin{figure}
\epsfxsize=4.truein
\centerline{\epsffile{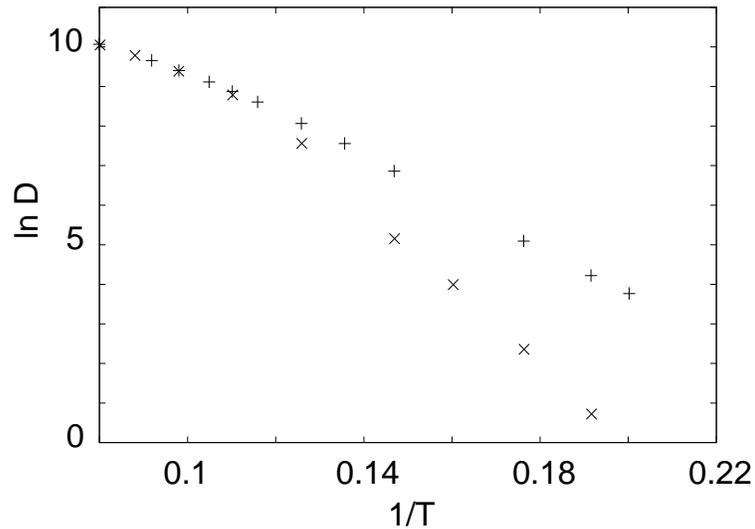}}
\caption{
Arrhenius plot ($\ln D$ versus $1/T$) for coverages
 $\theta = 1/2$ ($+$) and  $\theta = 2/3$ ($\times$).
\label{f4}}
\end{figure}

\begin{figure}
\epsfxsize=4.truein
\centerline{\epsffile{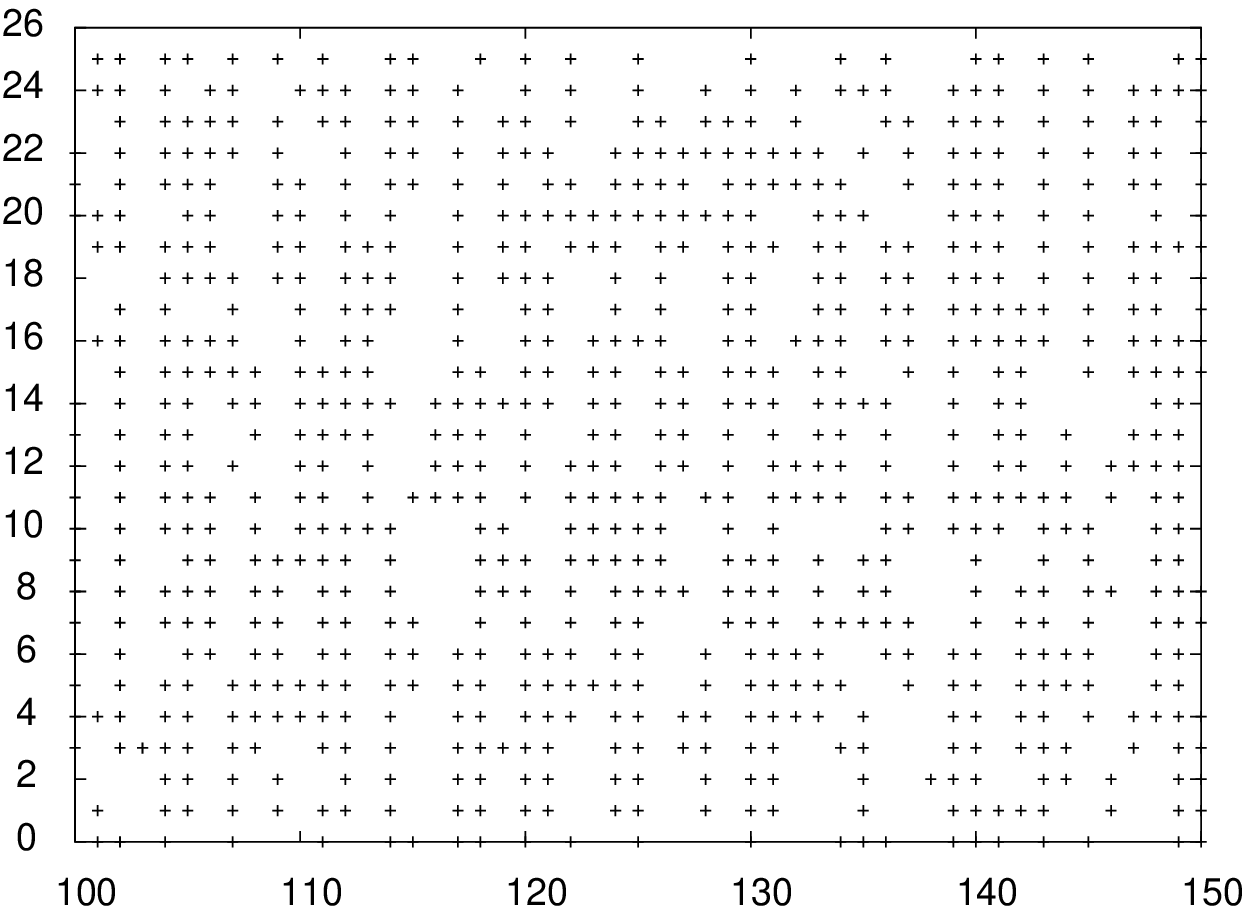}}
\vspace{0.5cm}
\caption{
Snapshot of the particle distribution in the $25\times 200$ system 
after  $10^6$ MC steps at $T = 11.35 J_0$. 
\label{f3}}
\end{figure}

\begin{figure}
\epsfxsize=4.truein
\centerline{\epsffile{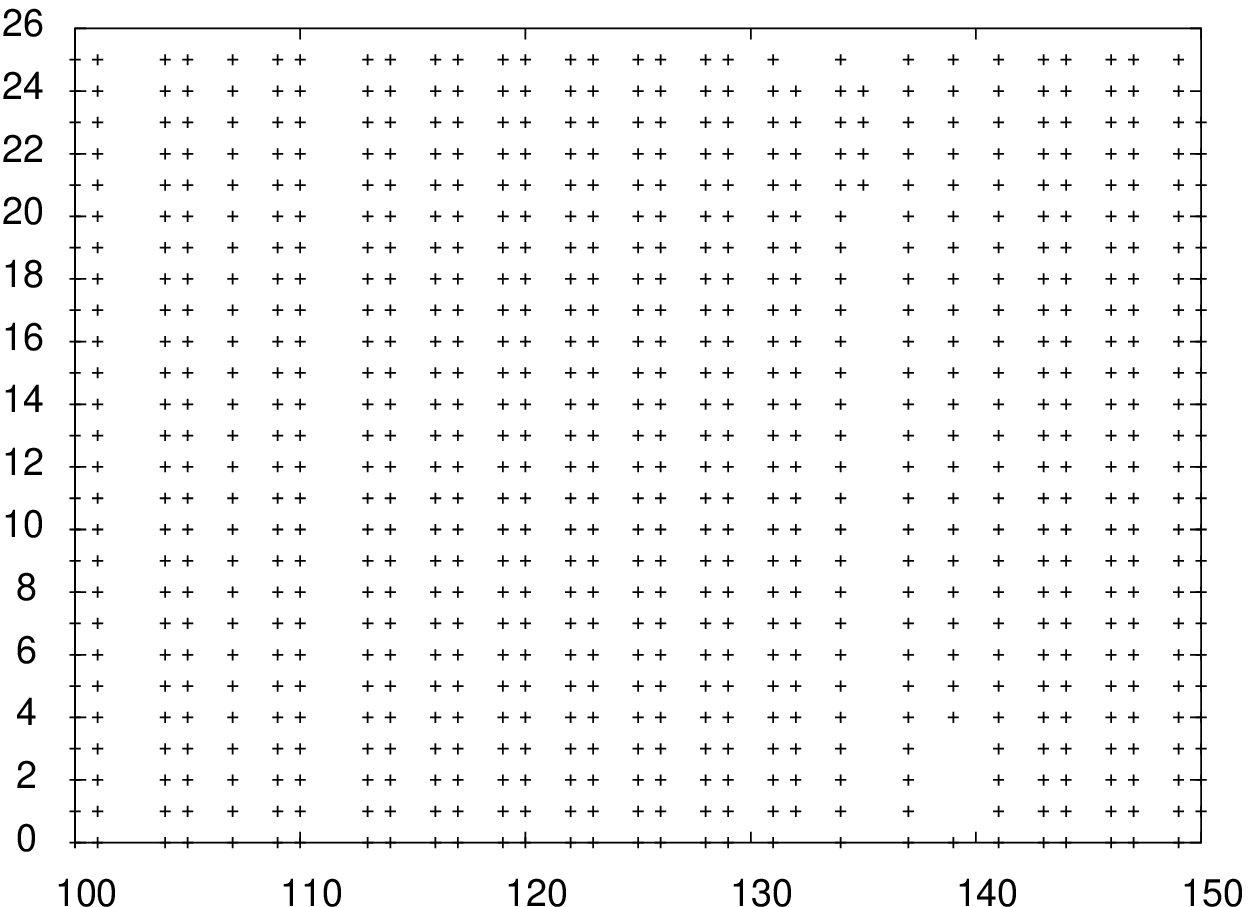}}
\vspace{0.5cm}
\caption{
Snapshot of the particle distribution in the $25\times 200$ system 
after  $10^6$ MC steps at $T = 4.55 J_0$. 
\label{f2}}
\end{figure}

We have simulated diffusive particle flow for coverages
corresponding to the structures  $p(1\times 2)$  ($\theta= 1/2$)  and
 $(1\times 3)-2X$ ($\theta= 2/3$) for the temperature range
 $ 4.55J_0 <T <12.5 J_0$ ( $ 1.18 T_{m,1/2} <T < 3.24 T_{m,1/2}$). 
As can be clearly seen from Fig.~\ref{f4}, which shows Arrhenius plots 
of the effective diffusion coefficient for the coverages 0.5 and 0.66, 
the effective diffusion coefficients are identical at the highest temperatures. 
Here the correlation radii in both directions are very small so that they do not 
play any important role. A snapshot of the situation for $\Theta = 1/2$ in a  
$25\times 200$ system 
after $10^6$ MC steps at 
$T=11.35 J_0 = 2.94 T_{m,1/2}$ is shown in Fig.~\ref{f3}, demonstrating the loss 
of correlations at this temperature. 
With decreasing temperature, however, the correlation radii increase, particularly 
along the chains, as just shown in Fig.~\ref{f11}. 
A snapshot, again for $\theta = 0.5$, after  $10^6$ MC steps at $T = 4.55 J_0$ 
($T = 1.18 T_{m,1/2}$), shown in Fig.~\ref{f2}, underlines this property. 
As is seen in  Fig.~\ref{f4} these correlations now result in different 
diffusion coefficients for the two coverages considered. The different slopes
of the Arrhenius plots in Fig.~\ref{f4} are evidence for the different
activation energies in these two cases.  
We feel that these simulations clearly demonstrate  
the close connection between  
the ordered low-temperature structures and the effective diffusion coefficient 
at temperatures well above the melting temperature. This connection is
established by the 
correlations of  rather long range
which remain 
far above the melting temperature in these strongly anisotropic systems. 

\subsection
{Diffusion and Point Defects.}
\label{def}
\begin{figure}
\epsfxsize=2.5truein
\centerline{\epsffile{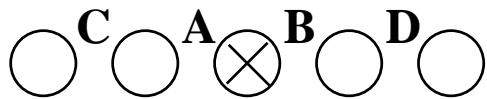}}
\vspace{0.5cm}
\caption{Atomic chain pinned by defect.$\times$ show the pinned atom.
\label{f70}}
\end{figure}

In order to get direct information about the main mechanism of diffusion, we 
have also studied diffusion in the presence of a small
concentration of  strong pinning centers of the random
field type, i.e. particles 
located at pinning centers have been fixed there. 
Due to the strong attractive interactions along the particle chains,  
correlation along the $Y$ axis remains strong well above
the transition temperature also in presence of the defects. 
For simplicity,
the same interaction parameters as for other particles 
have been used for these bound particles.  
This simplification can not change the physics we discuss 
in this article.
To clarify this point consider the chain of adatoms in Fig.~\ref{f70}.
If the pinned atom interacts with neighbors with 
the same interactions as unpinned atoms, 
the kink will appear 
by cutting the bond $A$ or $B$ on the chain 
and the activation energy will be  $ \vert J_{y}\vert $ 
(note that  $J_{y}<0 $).
Let us assume that the  pinned atom interacts 
with neighbors with interactions
$\vert J_{yd}\vert  \geq \vert J_{y}\vert  $. 
In the case   $\vert J_{yd}\vert  -\vert J_{y}\vert  \gg T$, 
the kink will appear 
by cutting the bond $C$ or $D$ on the chain. This means that 
the activation energy of the kink will remain 
the same as in the case   $J_{yd} = J_{y}$. 
but  the defect ``extends'' its size along the chain, i.e. 
the pre-exponential factor in the diffusion coefficient
will decrease. 
In the opposite case $\vert J_{yd}\vert  < \vert J_{y}\vert $ 
the chain can also not move through the defect atom which
is infinitely strongly bound with the substrate. This is  
true even in the case  $J_{yd}=0 $. The activation energy in 
this case remains   $\vert J_{y}\vert $ which is the kink energy
on the chain.  

The above discussion shows 
that activation energy will not change significantly by
varying strength of the defect interaction and the use of
the interaction  set of  non-pinned atoms is a reasonable
and convenient simplification. 
If diffusion is now limited mainly by the 
residence time of the chains at  the defect sites, we expect
that the activation energy is also not changed significantly
by varying  defect  concentrations, but a strong influence on the 
pre-exponential factor is expected.  
If the soliton mechanism is 
effective, Eq.~\ref{dif5} should hold as long as
the  defect concentration does not become too small
(see previous section).

\begin{figure}
\epsfxsize=4.truein
\centerline{\epsffile{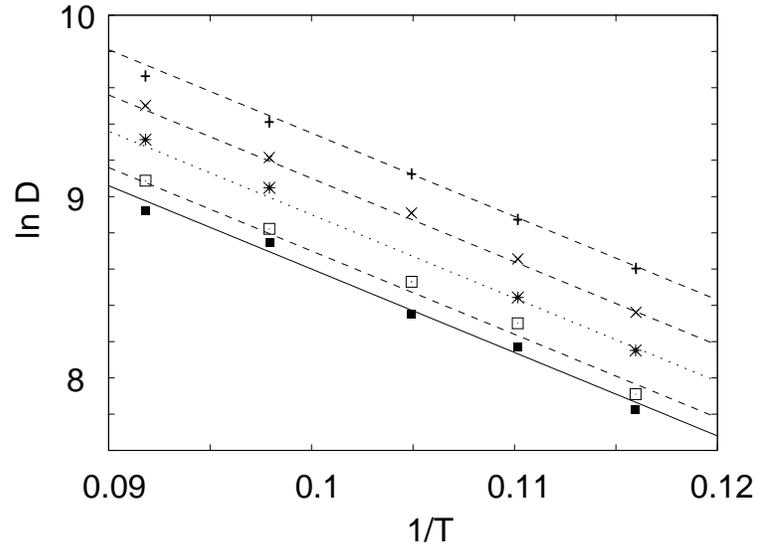}}
\vspace{0.5cm}
\caption{Arrhenius plot ($\ln D$ versus $1/T$) for coverage
 $\theta = 1/2$ without point defects($+$),
 with point defects at the following  concentrations:
 (1/64)($\times$),
 (1/32) (asterisks),
 (3/64) (open squares) and
 (1/16) (filled squares). Note that the lines have the same slope 
i.~e. the activation energy does not change.  
\label{f7}}
\end{figure}

The behavior just described 
is indeed found in our simulations, 
as Fig.~\ref{f7} demonstrates.
In this figure, Arrhenius plots of the diffusion coefficient 
are shown for a coverage $\theta = 1/2$ without point defects,
and  for concentrations of point defect varying between
(1/64) to (1/16). First, please note that already the smallest concentration 
of point defects (1/64) results in an observable
change of the diffusion coefficient. 
This is an important qualitative confirmation of the collective
nature of diffusion in our model because it directly contradicts 
the behavior expected for a 
single particle diffusion mechanism.  
If mass transport would happen 
mainly via single particle diffusion, 
these pinning centers would not be able to change significantly
the mass transport. Only for extended
mass carriers can these rare pinning centers 
have a strong effect on diffusion. 

Second, 
as expected from our qualitative arguments given above, 
the activation energy in presence of defects does not differ
from the case of the ideal substrate, and does not vary at 
different defect concentrations. To  show this more clearly, we have
fitted the simulation data by straight lines in 
Fig.~\ref{f7}. These lines are also included in this figure. 

\begin{figure}
\epsfxsize=5.truein
\centerline{\epsffile{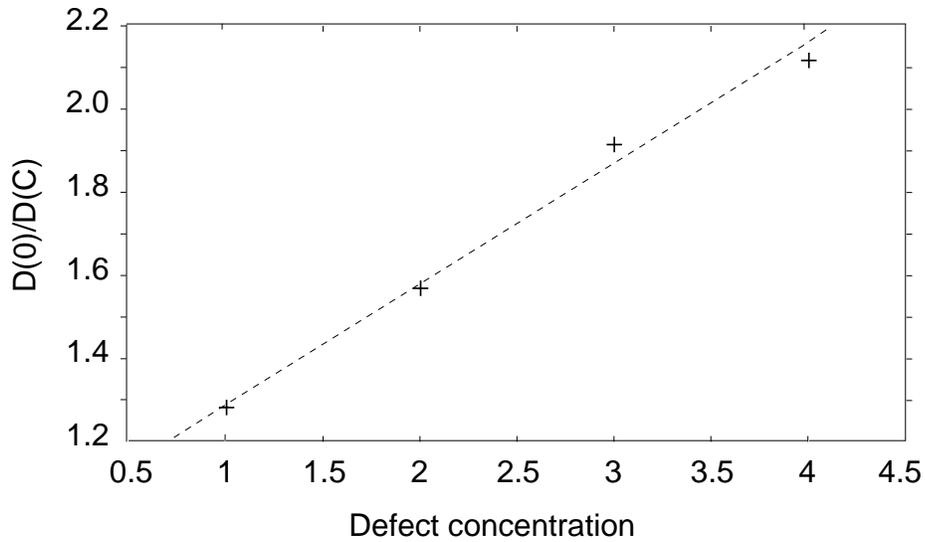}}
\vspace{0.5cm}
\caption{Ratio of the pre-exponential factors 
of the diffusion constants as function of point
defect concentration (in units 1/64)
following a linear dependence.
\label{f8}}
\end{figure}

Third, the pre-exponential factors strongly decrease with increasing 
defect concentration.  The resulting dependence of the 
ratio of the pre-exponential factors  $D(0)/D(\theta_d)$ 
of the diffusion constants as a function of the point-defect concentration 
is presented in Fig.~\ref{f8}.
The pre-exponential factor depends on $\theta_d$
as $D(\theta_d)=D(0)/(18.56\theta_d +1)$. 
The finite limit of  $D(\theta_d)$ when $\theta_d\to 0$
is due to the finite diffusion time between defects
which is comparable, at lowest defect density,
with the   
residence  time $\tau_{dep}$ on defects.
These results fully agree with our qualitative expectations outlined above.
Therefore, we think that  the results of our simulations can be regarded as a confirmation
of the soliton diffusion model.  

\section
{Conclusion.}
To conclude we have studied  the mass transfer 
in a system with highly anisotropic
interactions, which result in  chain structures at low temperature.
We have used both an analytical approach and numerical simulations. 
We propose that a mechanism of mass transfer similar to soliton diffusion
is effective even at temperatures much higher than the disordering (melting)
 temperature. 
Our computer simulations corroborate this proposal: 
for a given coverage we find a connection between the effective diffusion 
coefficient at
temperatures above the melting temperature and the phase that is stable at
this coverage for low temperatures.
An important test of this diffusion mechanism
is the  dependence of the effective diffusion coefficient on
concentration of frozen defects. 
For the case of defects which are deep traps we predict that  the activation 
energy depends 
weakly on the concentration of these  defects,
whereas the pre-exponential factor is inversely proportional to it.  
Although the 
influence of  frozen disorder should be most pronounced  for temperatures 
below the melting transition, it is still observable  
above this transition.
This prediction has also been confirmed by Monte-Carlo
simulations, which show 
that already a small concentration of
frozen point disorder strongly influences the 
diffusion coefficient due to the dominant mechanism of soliton-like diffusion. 
This strong influence of small concentrations of defects on the diffusion coefficient
would not be understandable for a predominant single particle diffusion mechanism.

The above results are not sensitive to the details of the adatom-adatom
interaction and to the microscopic model describing adatom jumps from one site
to another. This is both a strong and a weak point of our approach.
Although our general approach shows that this mechanism can be tested 
experimentally by introducing small 
concentrations of frozen defects, 
significantly more detailed knowledge is needed in order to be able to describe
a specific experimental system quantitatively.
However, our ``phenomenological'' model  can be used
much more generally for systems which have similar correlation properties,
but presumably totally different microscopic  mechanisms of mass transfer.
One example are vicinal surfaces. In this case we have strong 
correlations along the steps, relatively 
weak interaction between them and kinks as the elementary exitations,
which are responsible for 
step motion. However, the microscopic
mechanisms  of the kink motion on the step and on the adatom chain
are totally different. We plan to discuss the above similarity
elsewhere.

\section
{Acknowledgments}
We benefitted from discussions with  A.G.Naumovets.
The work is supported by the Nieders\"achsische Ministerium f\"ur 
Wissenschaft and Kultur and by the  Volkswagen Stiftung.
One of us (I.L.) was partly supported 
by the grants DE-FG03-96ER45598, NSF DMR-97-05182, 
THECB ARP 010366-003. 

\newpage

\end{document}